\documentclass[11pt,a4paper]{article}
\usepackage{jcappub}
\usepackage{amsmath,amssymb,graphics,epsfig,subfigure}
\usepackage{color}

\title{Curvature radius and Kerr black hole shadow}

\author{Shao-Wen Wei$^{1,2}$,}
\author{Yuan-Chuan Zou$^{3,4}$,}
\author{Yu-Xiao Liu$^{1}$,}
\author{Robert B. Mann$^{2}$}
\affiliation{$^{1}$Institute of Theoretical Physics $\&$ Research Center of Gravitation, Lanzhou University, Lanzhou 730000, People's Republic of China\\
$^{2}$Department of Physics and Astronomy, University of Waterloo, Waterloo, Ontario, Canada, N2L 3G1\\
$^{3}$School of Physics, Huazhong University of Science and Technology, Wuhan 430074, People's Republic of China\\
$^{4}$Perimeter Institute, 31 Caroline St., Waterloo, Ontario, N2L 2Y5, Canada}

\emailAdd{weishw@lzu.edu.cn, zouyc@hust.edu.cn, liuyx@lzu.edu.cn, rbmann@uwaterloo.ca}

\abstract{
We consider applications of the curvature radius of a Kerr black hole shadow and propose three new approaches to simultaneously determine the black hole spin and inclination angle of the observer. The first one uses only two symmetric characteristic points, i.e., the top and the bottom points of the shadow, and is the  smallest amount of data employed to extract information about spin and inclination angle amongst all current treatments. The second approach shows that only measuring the curvature radius at the characteristic points can also yield the black hole spin and the inclination angle. The observables used in the third approach have large changes to the spin and the inclination angle, which may give us a more accurate way to determine these parameters.  Moreover, by modeling the supermassive black hole M87* with a Kerr black hole, we calculate the angular size for these curvature radii of the shadow. Some novel properties are found and analyzed. The results may shine new light on the relationship between the curvature radius and the black hole shadow, and provide several different approaches to test the nature of the black hole through the shadow.}


\begin{document}


\maketitle

\section{Introduction}

 Very recently, the Event Horizon Telescope (EHT) has showcased the first image of the supermassive black hole M87* \cite{EHT,EHT2,EHT6}. This fruitful outcome reveals a fine structure near the black hole horizon. The photon ring and shadow have been clearly observed, which  provides us with a good opportunity to test general relativity  in the regime of strong gravity. Combined with   upcoming telescope surveys, such as the Next Generation Very Large Array \cite{Hughes2015} and the Thirty Meter Telescope \cite{Sanders2013}, more high-resolution observations will yield important new information about  strong gravity.

A black hole shadow is a two dimensional dark zone in the celestial sphere caused by the strong gravity of the black hole. It was first studied by Synge in 1966 for a Schwarzschild black hole \cite{Synge}. Later, a formula for the angular radius for the shadow was given by Luminet \cite{Luminet}. In general, the shadow cast by a non-rotating black hole is a standard circle, whereas for a rotating black hole the shadow will be elongated in the direction of the rotating axis  due to  spacetime dragging effects \cite{Bardeen,Chandrasekhar}.  In order to match  astronomical observations,
Hioki and Maeda \cite{Hioki} proposed two observables based on the characteristic points on the boundary of the Kerr shadow. One approximately describes the size of the shadow, and another describes the deformation of its shape from a reference circle. This approach has been extended to other black holes \cite{Johannsen,Ghasemi,Bambi,Amarilla,Stuchlik,Amarilla13,Nedkova,Wei,Tsukamoto,Bambi3,Atamurotov,Mann,WWei,FAtamurotov2,AmirAhmedov,Songbai,Balendra,Tsukamoto2,Jiliang,Rajibul,hou,Cunha,Cunha2,Cunha3,Tsupko,Perlick,Kocherlakota,WangXu,WeiLiu,Younsi,cBambi,Akashaa}, and a coordinate-independent study of  the shadow by making use of Legendre polynomials has been carried out \cite{Abdujabbarov}.

Information about the properties of a black hole is contained on the boundary of  its shadow. So if the boundary curve is uniquely determined, we can extract this information from it. Motivated by this idea we recently introduced a new concept, the local curvature radius \cite{WeiLiu}, from the viewpoint of   differential geometry. For each fixed black hole spin and  inclination angle of the observer, we found there exists one minimum and one maximum of the curvature radius upon  taking the symmetry of the black hole shadow into account.  Employing this property we showed that both the spin and the inclination angle can be uniquely determined.

We further extended our application of the local curvature radius to construct a topological quantity \cite{WeiLiu} associated with the shadow. The value of this quantity is unity for an arbitrary Kerr black hole, but smaller than one for a naked singularity. Using this quantity, we can therefore distinguish a black hole from a naked singularity. It can be used to measure the topological structure for the multi-shadows in some special spacetimes.

The aim of this paper is to further explore  applications of the local curvature radius. Noting that  the maximum curvature always occurs at the left point of the shadow whereas the corresponding point of the minimum curvature varies with the black hole parameters \cite{WeiLiu}, we focus on  how to extract black hole parameters  from this information. Rather than measuring the curvature radius at every point on the shadow boundary, we can exploit the symmetry of the shadow to obtain its characteristic points, making use of a detailed study we have previously carried out \cite{WeiP}.
In the Kerr spacetime, these points have analytical coordinates for the equatorial observer, and  away from the equatorial plane these points are also very easy to work out.

It is therefore natural to combine the curvature radius and characteristic points to determine the black hole parameters. Three novel approaches are presented in this paper. These results demonstrate that the local curvature radius is a very useful tool in the  study of black hole shadows. We apply our results on the curvature radius to the supermassive  black hole M87*.

Our paper is organized as follows. In Sec. \ref{null}, we briefly review the null geodesics and shadow for the Kerr black hole. Then the curvature radius is introduced. In Sec. \ref{angle}, we show how to determine the black hole spin and the inclination angle of the observer by making use the curvature radius. Three approaches are proposed for different purposes. We then consider the supermassive M87* black hole in Sec. \ref{Application}. Finally, the conclusions and discussions are presented in Sec. \ref{Conclusion}.

\section{Null geodesics and shadow}
\label{null}

Here we provide a brief review of the null geodesics and shapes of the shadow in a Kerr spacetime.

In Boyer-Lindquist coordinates, the line element in a Kerr spacetime is
\begin{eqnarray}
 ds^{2}=-\left(1-\frac{2Mr}{\rho^{2}}\right)dt^{2}+\frac{\rho^{2}}{\Delta}dr^{2}+\rho^{2}d\theta^{2}
 -\frac{4Mra\sin^{2}\theta}{\rho^{2}}dtd\phi+\frac{((r^{2}+a^{2})^{2}-\Delta a^{2}\sin^{2}\theta)\sin^{2}\theta}{\rho^{2}}d\phi^{2},\label{metric}
\end{eqnarray}
where the metric functions are
\begin{eqnarray}
 \Delta=r^{2}-2Mr+a^{2},\quad
 \rho^{2}=r^{2}+a^{2}\cos^{2}\theta.
\end{eqnarray}
The parameters $M$ and $a$ are, respectively, the black hole mass and spin. The horizons in a Kerr spacetime can be obtained by solving $\Delta(r)=0$, which are located at
\begin{eqnarray}
 r_{\pm}=M\pm \sqrt{M^{2}-a^{2}}.
\end{eqnarray}
There are two horizons for $|a|<M$ and one horizon for $|a|=M$; no horizon exists for $|a|>M$, which corresponds to a naked singularity.

\begin{figure}
\center{\includegraphics[width=6cm]{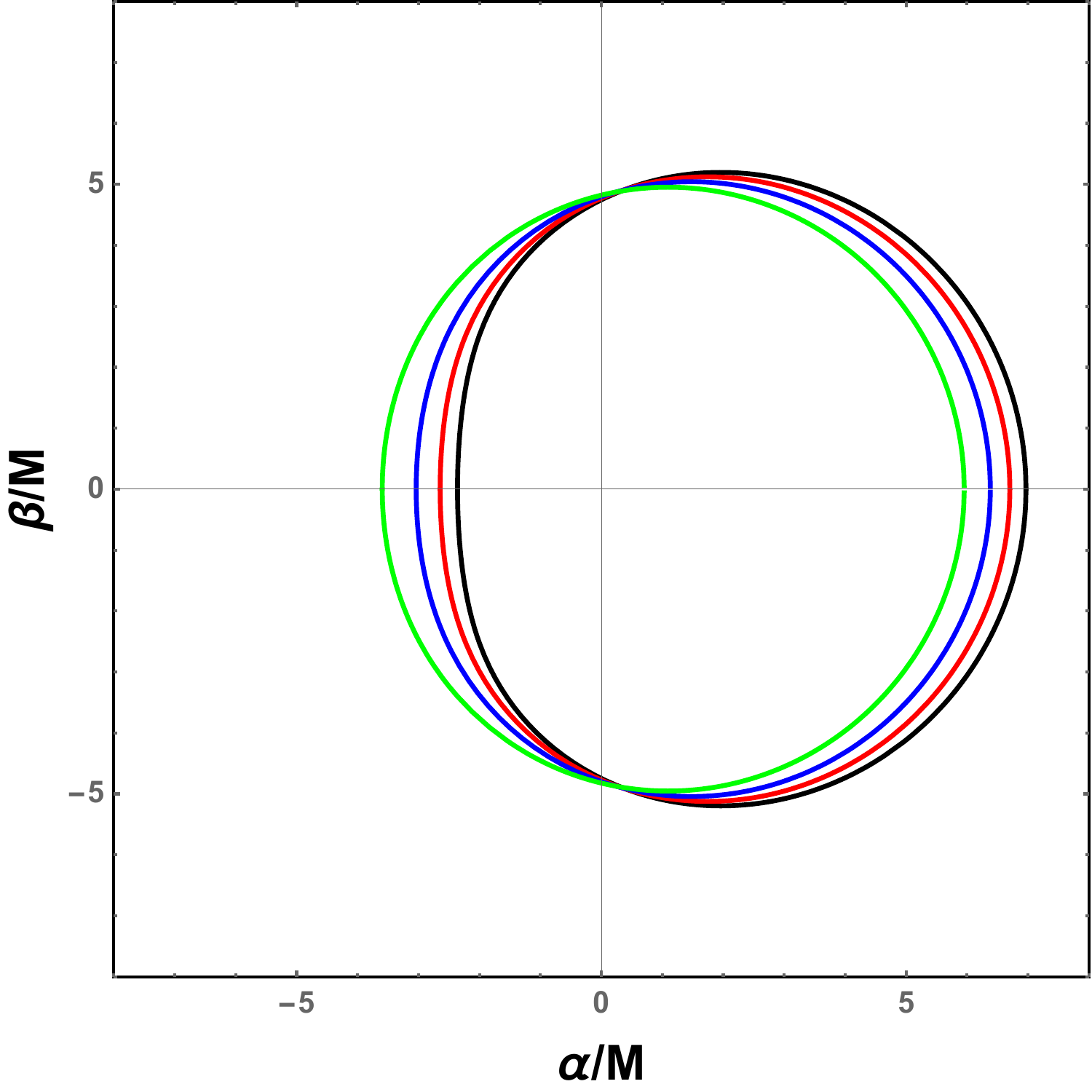}}
\caption{Shapes of the shadow in Kerr spacetime for $a/M=0.98$ with the inclination angle $\theta_{0}=\frac{\pi}{2}, \frac{\pi}{3}, \frac{\pi}{4}$, and $\frac{\pi}{6}$ from right to left.}\label{ppTNss}
\end{figure}

The null geodesics in the Kerr spacetime are given by the solutions to the equations
\begin{eqnarray}
 \rho^{2}\frac{dt}{d\lambda}&=&a(l-aE\sin^{2}\theta)
       +\frac{r^{2}+a^{2}}{\Delta}\Big(E(r^{2}+a^{2})-al\Big),\\
 \rho^{2}\frac{dr}{d\lambda}&=&\sqrt{\Re},\label{Rad}\\
 \rho^{2}\frac{d\theta}{d\lambda}&=&\sqrt{\Theta},\\
 \rho^{2}\frac{d\phi}{d\lambda}
     &=&(l\csc^{2}\theta-aE)+\frac{a}{\Delta}\Big(E(r^{2}+a^{2})-al\Big),\label{phiequation}\label{thetaequation}
\end{eqnarray}
where $\lambda$ is the affine parameter. The functions ${\Re}$ and $\Theta$ are given by
\begin{eqnarray}
 {\Re}&=&\left(a^2 E-al+E r^2\right)^2-\Delta
   \left(\mathcal{Q}+(l-a E)^2\right),\\
 \Theta&=&\mathcal{Q}-(l\csc\theta-a
   E \sin\theta)^2+(l-a E)^2.
\end{eqnarray}
Here the conserved quantities $l$ and $E$ are the angular momentum and energy of the test particle, which are related with the Killing vector fields $\partial_{\phi}$ and $\partial_{t}$, respectively. The conserved Carter constant $\mathcal{Q}$ is related to the Killing-Yano tensor field in the Kerr spacetime \cite{Yano,Penroser}. Moreover, one can introduce two new parameters $\xi$ and $\eta$
\begin{eqnarray}
 \xi=\frac{l}{E}, \quad \eta=\frac{\mathcal{Q}}{E^{2}}.
\end{eqnarray}
By making use of the null geodesics, we can obtain the two celestial coordinates $\alpha$ and $\beta$, which are used to describe the shape of the shadow that an observer seen in the sky. For an observer of the inclination angle $\theta_{0}$, the celestial coordinates are given by
\begin{eqnarray}
 \alpha&=&-\xi\csc\theta_{0},\label{oe}\\
 \beta&=&\pm\sqrt{\eta+a^{2}\cos^{2}\theta_{0}-\xi^{2}\cot^{2}\theta_{0}}.\label{oe2}
\end{eqnarray}
On the boundary, the parameters $\xi$ and $\eta$ are given by
\cite{Young}
\begin{eqnarray}
 \xi&=&\frac{(3M-r_{0})r_{0}^{2}-a^{2}(M+r_{0})}{a(r_{0}-M)},\label{xx}\\
 \eta&=&\frac{r_{0}^{3}(4a^{2}M-r_{0}(3M-r_{0})^{2})}{a^{2}(r_{0}-M)^{2}}.\label{ee}
\end{eqnarray}
where $r_{0}$ varies between the light ring radii of direct and retrograde photons along the boundary of the shadow. This also provides us with a parametrization of the form of the shadow, and based on it,  some analytic results can be obtained Ref. \cite{WeiP}. For convenience, we show the shapes of the shadow for the black hole in Fig. \ref{ppTNss} with spin $a/M=0.98$, and  inclination angle $\theta_{0}=\frac{\pi}{2}, \frac{\pi}{3}, \frac{\pi}{4}$, and $\frac{\pi}{6}$ from right to left. In order to study them, we  introduced a new quantity,  the curvature radius $R$  \cite{WeiLiu}. From the viewpoint of differential geometry, if we measure the curvature radius at each point of the boundary, the shadow will be uniquely determined, and then we can exactly read out the black hole parameters in a given spacetime, or to distinguish different gravity theories via the structure of the shadow.

We now  briefly review how to calculate the local curvature radius  \cite{WeiLiu}. From Eqs. (\ref{oe}) and (\ref{oe2}), one can find that the celestial coordinates ($\alpha$, $\beta$) for each point on the boundary of the shadow are parametrized by $r_{0}$. Thus each value of $r_{0}$ corresponds to one point. In order to obtain the curvature radius, we consider three values of $r_{0}$, i.e., $r_{0}-\epsilon$, $r_{0}$, and $r_{0}+\epsilon$, yielding three points on the boundary, $\left(\alpha(r_{0}-\epsilon), \beta(r_{0}-\epsilon)\right)$, $\left(\alpha(r_{0}), \beta(r_{0})\right)$, and $\left(\alpha(r_{0}+\epsilon), \beta(r_{0}+\epsilon)\right)$ -- from these we can uniquely plot a circle of radius $R(r_{0}, \epsilon)$. In the limit $\epsilon\rightarrow0$, these three points approach the same  point, and $R(r_{0}, \epsilon)\to R(r_{0})$, which is just the curvature radius of the point $\left(\alpha(r_{0}), \beta(r_{0})\right)$. Adopting this method, we obtained the curvature radius for the Kerr black hole shadow, which reads \cite{WeiLiu}
\begin{eqnarray}
 R=\frac{64M^{1/2}(r_{0}^{3}-a^{2}r_{0}\cos^{2}\theta_{0})^{3/2}\left[r_{0}(r_{0}^{2}-3Mr_{0}+3M^{2})-a^{2}M^{2}\right]}
   {(r_{0}-M)^{3}\left[3(8r_{0}^{4}-a^{4}-8a^{2}r_{0}^{2})-4a^{2}(6r_{0}^{2}+a^{2})\cos(2\theta_{0})
    -a^{4}\cos(4\theta_{0})\right]}.\label{radiuscur}
\end{eqnarray}

Alternatively, we can parametrize $R$ with the length parameter $\lambda$ of the shadow. From this we proposed  \cite{WeiLiu} a topological quantity $\delta=\int \frac{d\lambda}{R(\lambda)}+\sum_{i}\theta_{i}$ to measure the topological structure of the shadow. It can also be used to distinguish a shadow cast by a black hole from that of a naked singularity. In fact, this intrinsic curvature radius is critical for testing the black hole parameter through the shadow, and we will show it in the next section.

\section{Determining spin and inclination angle}
\label{angle}

In this section, we   show how to determine the spin $a$ and inclination angle $\theta_{0}$ for the Kerr black hole by making use of the curvature radius.

\subsection{Approach I}

Generally, there are two ways to test the black hole parameters and the viewing angle of the observer. The first one is that we can completely determine every point on the boundary of the shadow, and then fit the theoretical model. Thus we can find out the best values for the black hole parameters in the expected spacetime. This method  can also be used to determine the possible modified theory. As shown above the shadow can be uniquely described by the curvature radius (\ref{radiuscur}). However, this task is extremely hard  because there are so many points on the boundary of the shadow. In order to make this task feasible we must reduce the number of the points used. This is the motivation of the second way.

\begin{figure}
\center{\subfigure[]{
\includegraphics[width=8cm]{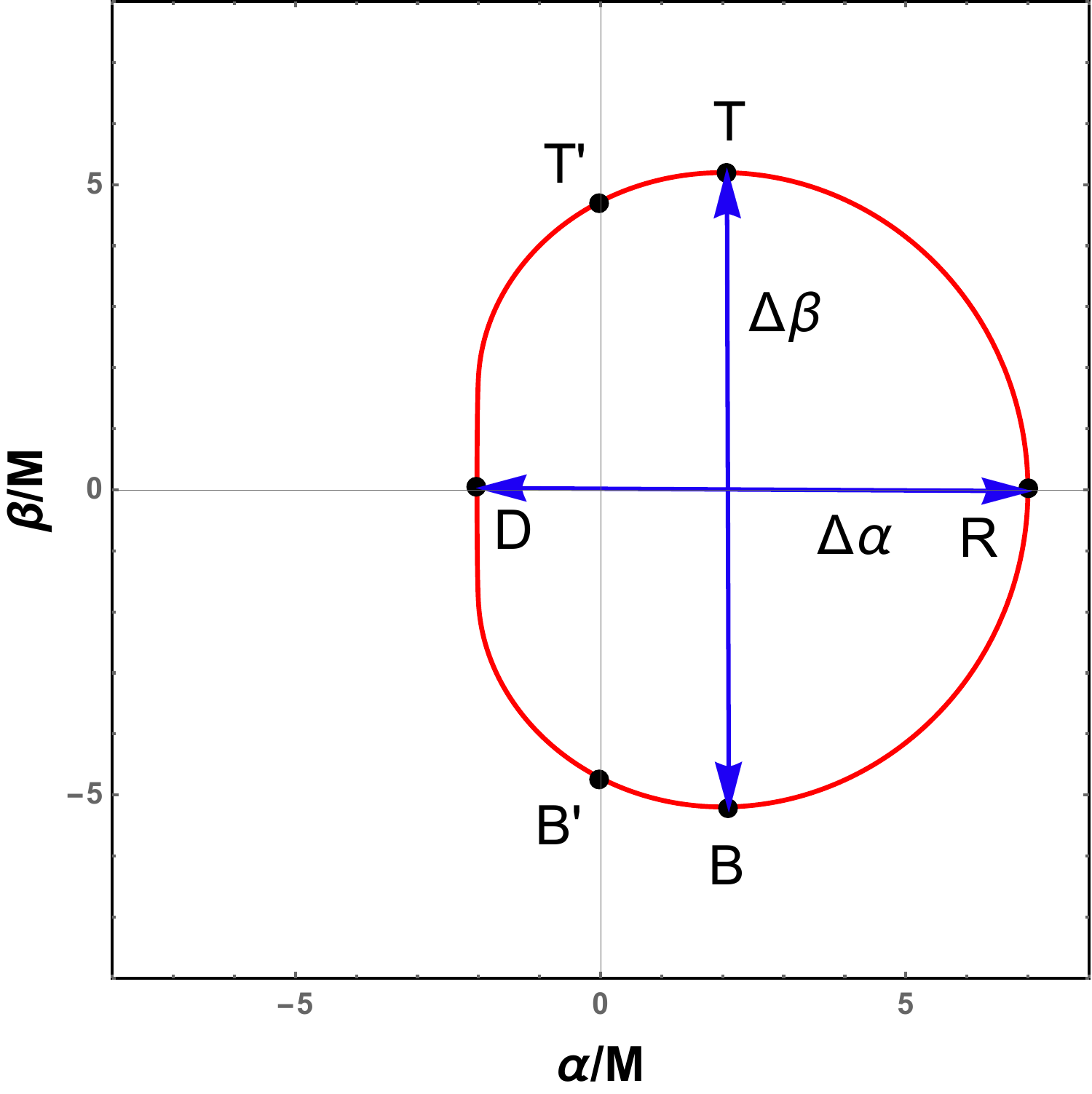}}}
\caption{Sketch picture for the characteristic points of a Kerr black hole shadow. The red circle denotes the shape of the shadow with a certain spin $a$ and viewing angle $\theta_{0}$. Characteristic points D, R, T, B are, respectively, the left, right, top, and bottom points of the shadow. Points T$'$ and B$'$ corresponds to $\alpha$=0. $\Delta\alpha$ and $\Delta\beta$ denote the horizontal and vertical diameters of the shadow.}\label{ppAB}
\end{figure}

For a Kerr black hole, if the inclination angle of the observer is known, one can measure any one of the observables discussed in Ref. \cite{WeiP} to obtain the black hole spin. However, if the angle is unknown, at least two observables must be included. We focus our attention on the characteristic points, since almost all astronomical observables are constructed from them.

For clarity, we show a schematic plot for these points in Fig. \ref{ppAB}. Characteristic points D, R, T, B are, respectively, the left, right, top, and bottom points of the shadow. Points T$'$ and B$'$ correspond to $\alpha$=0. These characteristic points were examined in detail in Ref. \cite{WeiP}.  In particular, the authors of Ref. \cite{Hioki} proposed two observables. One is the radius of the reference circle that passes through the points `T', `R', and `B'. Although the points `T' and `B' are symmetric, one must use them to find  the $\alpha$-axis, so three points are used to obtain the radius.  The other one is the distortion, which is defined by the ratio between the deformation and the radius of the reference circle. For this observable, four points `T', `B', `R', and `D' are used. After obtaining these two observables, one can get the Kerr black hole spin and the inclination angle of the observer. This method has been extended to other black hole backgrounds. So employing this method, four characteristic points on the shadow must be given in order to get $a$ and $\theta_{0}$. However, we still want to ask whether the number of the points can be further reduced.


When we examine the property of the curvature radius $R$ for the Kerr black hole, we find that both $a$ and $\theta_{0}$ can be determined by two symmetric characteristic points, the top point `T' and the bottom point `B' of the shadow. We summarize the method in the following.

Since this method closely depends on the points `T' and `B', let us start with them. The points are determined by
\begin{eqnarray}
 (\partial_{\alpha}\beta)_{a, \theta_{0}}=0.
\end{eqnarray}
Adopting the parametric forms of $\alpha$ and $\beta$, the condition will be reduced to
\begin{eqnarray}
 (\partial_{r_{0}}\alpha)^{-1}_{a, \theta_{0}}=0,\label{eq1}
\end{eqnarray}
or,
\begin{eqnarray}
 (\partial_{r_{0}}\beta)_{a, \theta_{0}}=0.\label{eq2}
\end{eqnarray}
The first condition (\ref{eq1}) gives
\begin{eqnarray}
 a(M-r)^{2}\sin\theta_{0}=0.
\end{eqnarray}
For $\theta_{0}\neq0$, one has $r=M$, which is smaller than the radius of the event horizon of a non-extremal black hole, and thus we turn to another condition. The second condition (\ref{eq2}) gives
\begin{eqnarray}
 r_{0}^{3}-3M r_{0}^{2}+3M^{2}r_{0}-a^{2}M=0,\label{eq3}
\end{eqnarray}
or,
\begin{eqnarray}
 r_{0}^{3}-3Mr_{0}^{2}+a^{2}\cos^{2}\theta_{0}r_{0}+a^{2}M\cos^{2}\theta_{0}=0.\label{eq4}
\end{eqnarray}

\begin{figure}
\center{\subfigure[]{\label{TBBath}
\includegraphics[width=8cm]{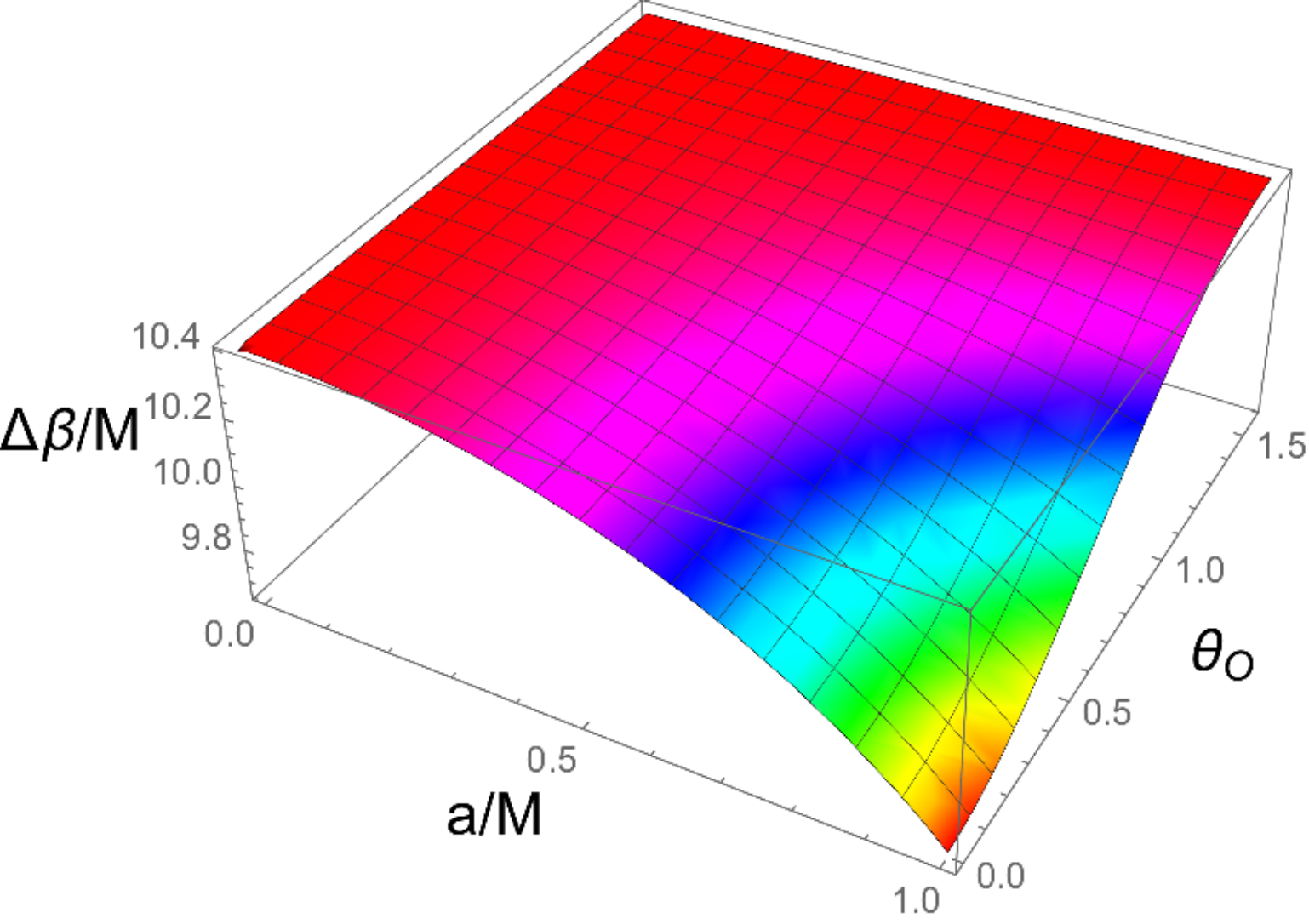}}
\subfigure[]{\label{TRTaTh}
\includegraphics[width=8cm]{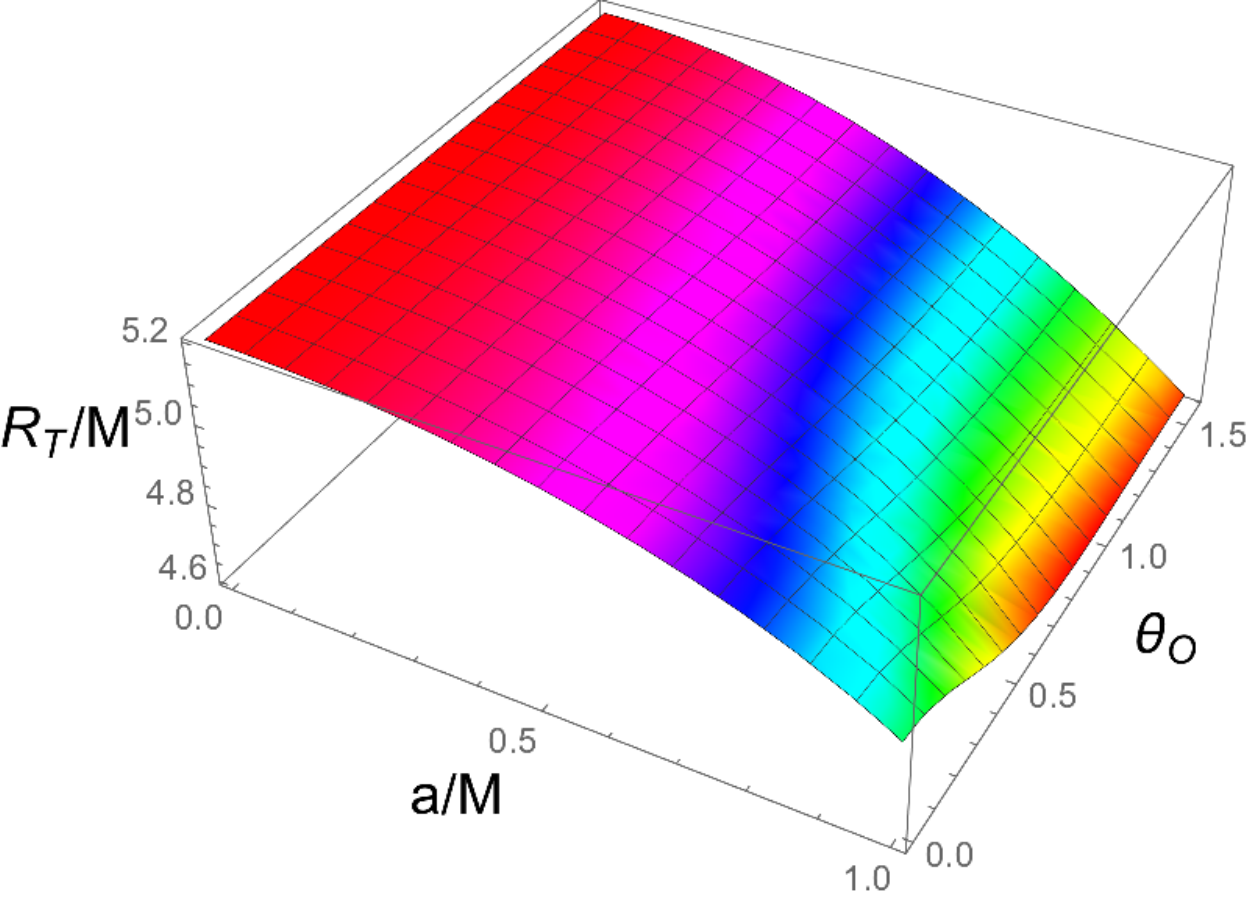}}}
\caption{(a) The behavior of the vertical diameter $\Delta\beta$ of the shadow as a function of $a$ and $\theta_{0}$. (b) The curvature radius $R_{\rm T}$ of top point `T' as a function of $a$ and $\theta_{0}$.}\label{ppBBath}
\end{figure}

Equation (\ref{eq3}) gives  a real root $r_{0}=M-\left(M^{3}-Ma^{2}\right)^{\frac{1}{3}}$, which is also smaller than the radius of the event horizon, so we abandon it. Finally, by solving Eq. (\ref{eq4}), we obtain
\begin{eqnarray}
 r_{0}=M+\frac{2}{\sqrt{3}}\sqrt{3M^2-a^{2}\cos^{2}\theta_{0}}
 \cos\left(\frac{1}{3}\arccos\left(\frac{3\sqrt{3}M(M^2-a^{2}\cos^{2}\theta_{0}^{2})}{(3M^2-a^{2}\cos^{2}\theta_{0})^{3/2}}\right)\right).
\end{eqnarray}
Inserting this result into ($\alpha$, $\beta$), one will obtain the coordinates ($\alpha_T$, $\beta_T$) of the point `T'.  The result is complicated and we will not show the expression. But from it we can extract the vertical diameter $\Delta\beta$ of the shadow
\begin{eqnarray}
 \Delta\beta=2 \beta_{\rm T},
\end{eqnarray}
which is plotted as a function of $a$ and $\theta_{0}$ in Fig. \ref{TBBath}. From it, we can find that $\Delta\beta$ decreases quickly when $a/M$ approaches 1 and $\theta_{0}$ approaches 0.

Furthermore, when the observer is located on the equatorial plane $\theta_{0}=\frac{\pi}{2}$, we have
\begin{eqnarray}
 \Delta\beta=6\sqrt{3}M,
\end{eqnarray}
which is independent of the black hole spin $a$. Alternatively, when $\theta_{0}=0$ and $a/M=1$, $\Delta\beta/M=4\sqrt{3+2\sqrt{2}}\approx9.6568$.

We next depict the curvature radius given in (\ref{radiuscur}) at point `T'  in Fig. \ref{TRTaTh}. For fixed $\theta_{0}$, $R_{\rm T}$ always decreases with the spin $a$. Interestingly, for the Schwarzschild black hole with $a/M=0$, $R_{\rm T}/M=3\sqrt{3}$ is independent of the inclination angle $\theta_{0}$, a consequence of the spherical symmetry of  the Schwarzschild black hole.

\begin{figure}
\center{\subfigure[]{
\includegraphics[width=8cm]{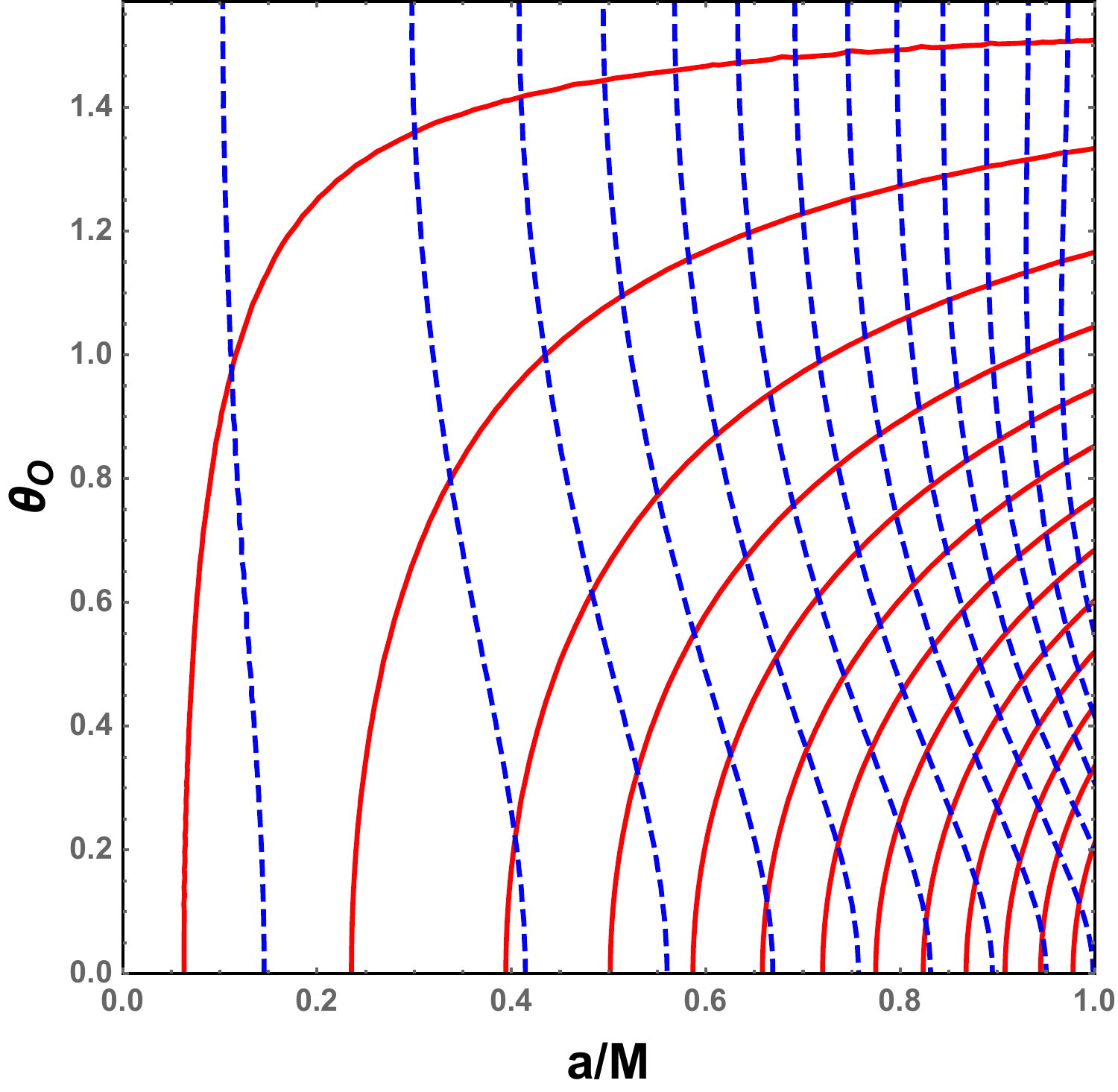}}}
\caption{Contour lines of $\Delta\beta$ and $R_{\rm T}$. The red solid lines are for $\Delta\beta/M$= 9.70, 9.76, 9.82, 9.88, 9.94, 10.00, 10.06, 10.12, 10.18, 10.24, 10.3, 10.36, 10.39 from bottom right to top left. The blue dashed lines are for $R_{\rm T}/M$= 4.65, 4.695, 4.74, 4.785, 4.83, 4.875, 4.92, 4.965, 5.01, 5.055, 5.1, 5.145, 5.19 from right to left.}\label{ppContr}
\end{figure}

If we have the shape of a Kerr black hole shadow from   astronomical observations, it is quite easy to obtain the $\alpha$-axis, according to the $\mathcal{Z}_{2}$ symmetry of the shadow, where points `T' and `B' can be used. However, it is impossible to obtain the $\beta$-axis without any other information. Nevertheless, we do measure the vertical diameter $\Delta\beta$ of the shadow. So only giving the shadow, we can determine the top point `T' and then get $\Delta\beta$. Moreover, the curvature radius $R_{\rm T}$ at point `T' can also be obtained by the geometric method. Thus we have $\Delta\beta$ and $R_{\rm T}$ from the shape, and both depend on the black hole spin $a$ and the inclination angle $\theta_{0}$. Next, we can solve $a$ and $\theta_{0}$ from these two observables. For an example, we plot the contour lines of $\Delta\beta$ and $R_{\rm T}$ in Fig. \ref{ppContr} in the $\theta_{0}$-$a$ plane with $\Delta\beta/M$= 9.70, 9.76, 9.82, 9.88, 9.94, 10.00, 10.06, 10.12, 10.18, 10.24, 10.3, 10.36, 10.39 from bottom right to top left and $R_{\rm T}/M$= 4.65, 4.695, 4.74, 4.785, 4.83, 4.875, 4.92, 4.965, 5.01, 5.055, 5.1, 5.145, 5.19 from right to left. The red solid lines are for $\Delta\beta$ and blue dashed lines for $R_{\rm T}$. From this figure, we can directly read out the black hole spin $a$ and the location of the observer $\theta_{0}$ with given $\beta_{\rm T}$ and $R_{\rm T}$.

A few comments are worth making at this point about this method.   i) It uses only  two symmetric characteristic points to simultaneously determine both $a$ and $\theta_{0}$.  ii) It depends only on the intrinsic properties of the shape of the shadow, and is independent of the coordinates.  iii) Although more black hole parameters are included when considering other non-Kerr black holes, we believe that   by making use of the curvature radius,   the number of the points that must be used will be greatly reduced.

\subsection{Approach II}

From approach I, we see that the parameters $a$ and $\theta_{0}$ can be both determined by measuring $\beta_{\rm T}$ and the curvature radius at point `T'. Can we determine $a$ and $\theta_{0}$  using only the curvature radius?   The answer is `yes' as we shall now demonstrate.

We have previously shown \cite{WeiLiu} that when excluding the $\mathcal{Z}_{2}$ symmetry, there exists one maximum and one minimum of the curvature $R_{\rm T}$ for fixed $a$ and $\theta_{0}$. Combining these two curvature radii, we show that $a$ and $\theta_{0}$ can be uniquely determined \cite{WeiLiu}. However, although the maximum one is at point `D', the minimum is uncertain.  But we can replace the point of   minimum curvature with other characteristic points, for example point `R' or `D'. Thus, we need only measure the curvature of any two points of `T', `R', and `D' to obtain $a$ and $\theta_{0}$.  According to this idea, by measuring the curvature for any two points without symmetry on the boundary of the black hole shadow, we can obtain $a$ and $\theta_{0}$. Nevertheless, one should keep in mind that, given a shape of the shadow, we must first use points `T' and `B' to find out the $\alpha$-axis. So in this approach, at least, three points must be included in. Thus comparing with approach I, this approach uses more points. Hence, it also provides us a novel approach for using the curvature radius.

\begin{figure}
\center{\subfigure[]{\label{TLRa}
\includegraphics[width=5cm]{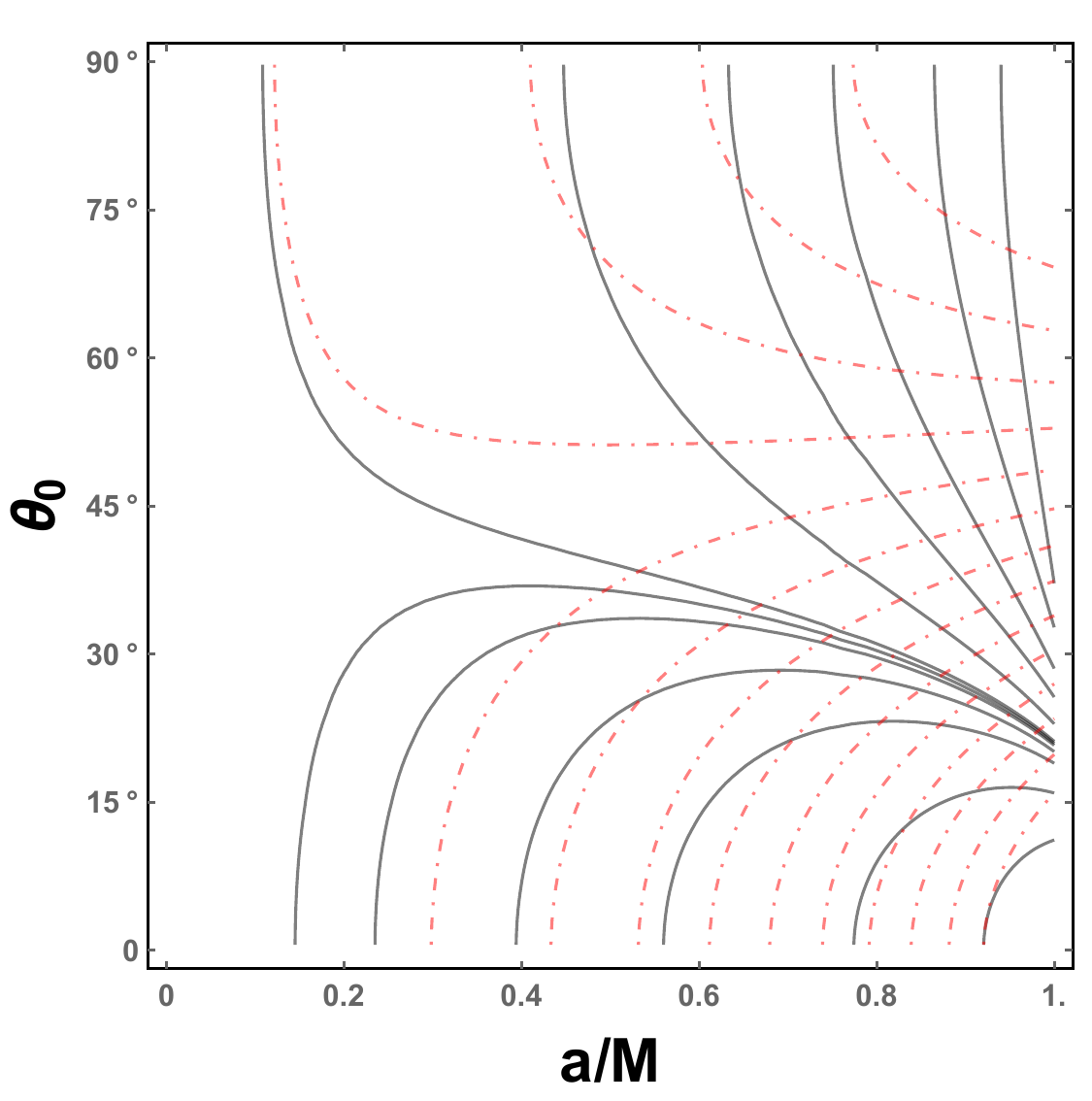}}
\subfigure[]{\label{TLTb}
\includegraphics[width=5cm]{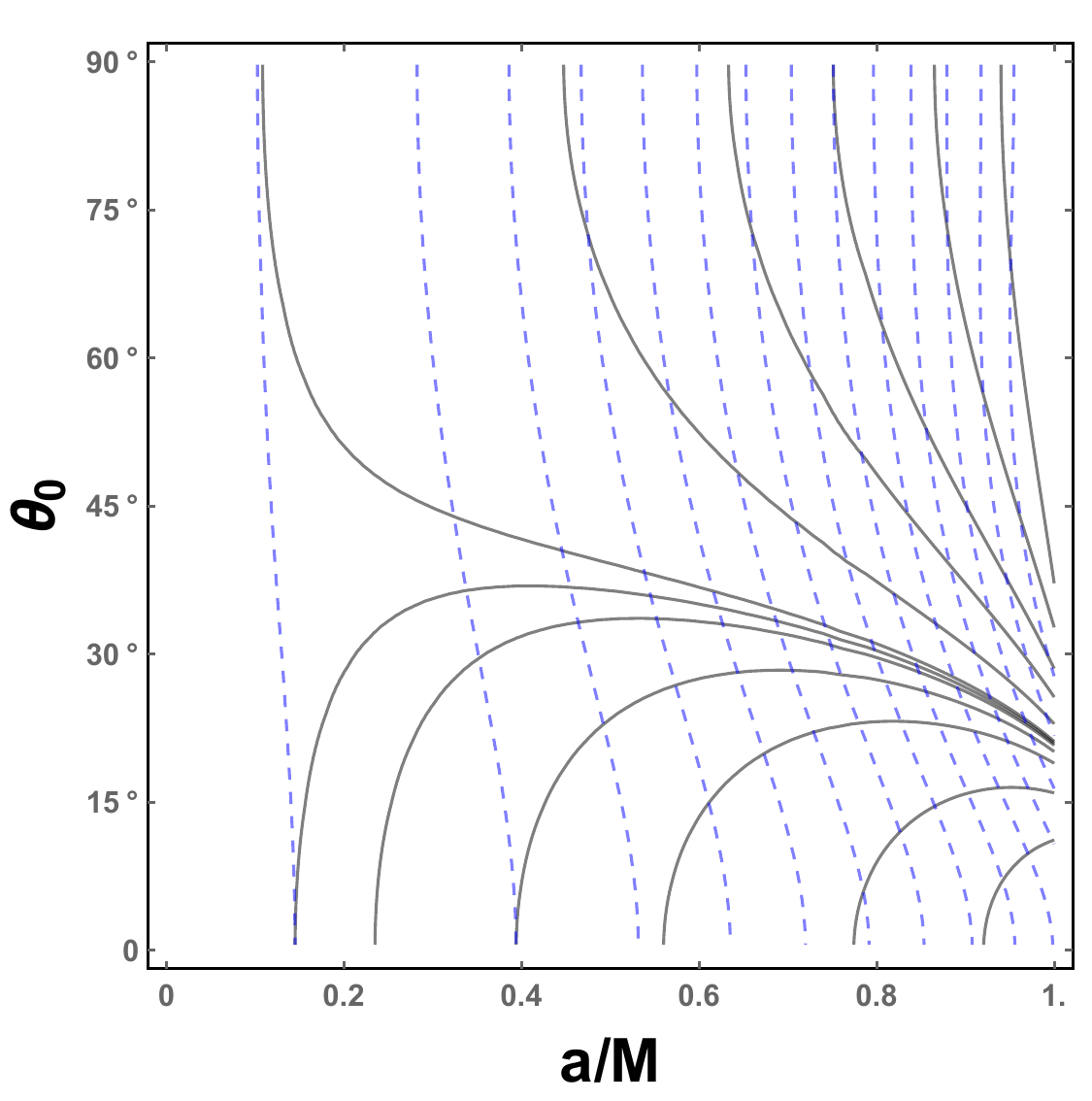}}
\subfigure[]{\label{TRTc}
\includegraphics[width=5cm]{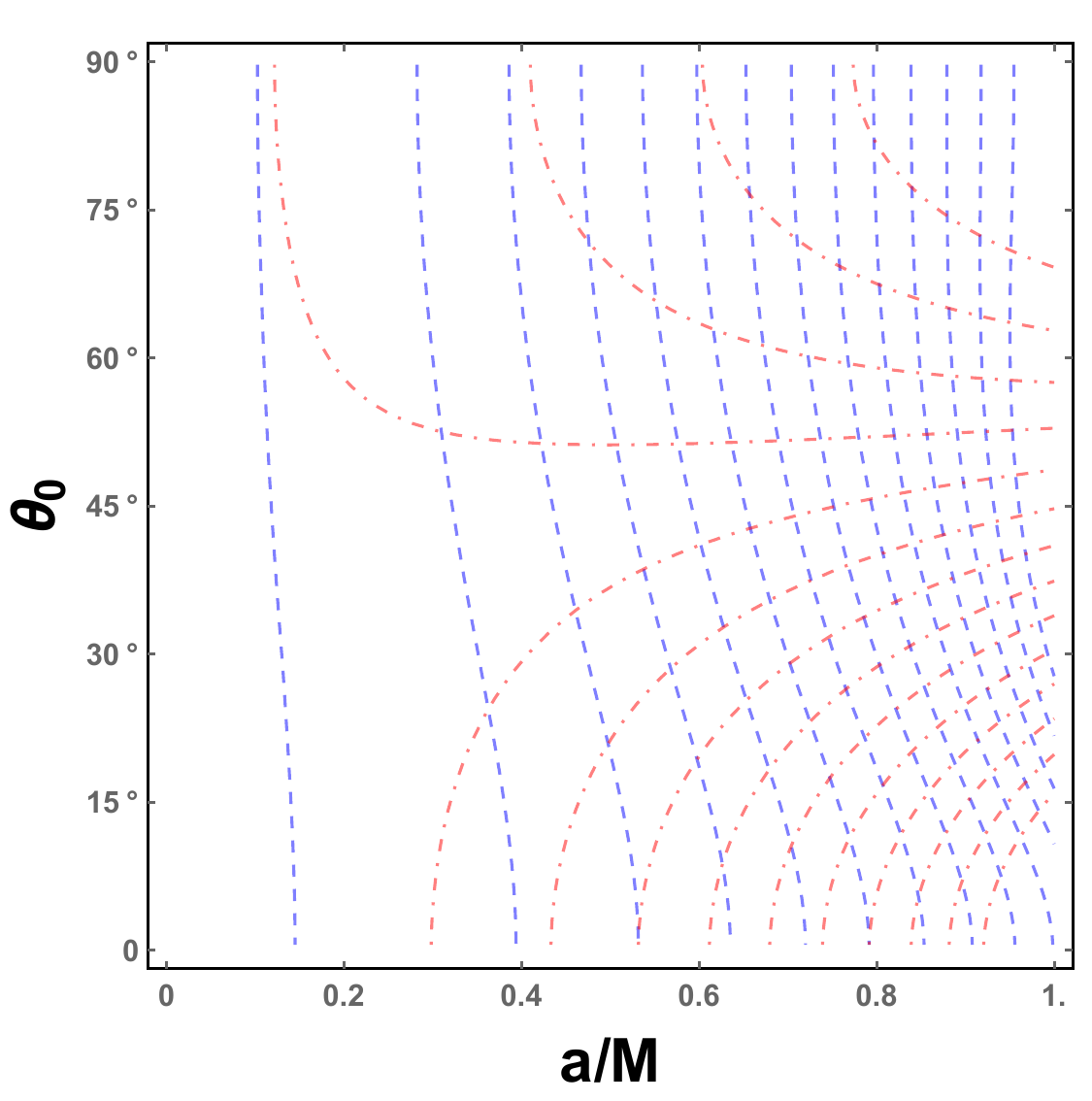}}}
\caption{Contour lines of $R_{\rm D}$, $R_{\rm T}$, and $R_{\rm R}$ in $\theta_{0}$-$a$ plane. The black solid lines are for $R_{\rm D}/M$=4.9, 5, 5.1, 5.15, 5.18, 5.19, 5.2, 5.3, 5.5, 5.8, 6.5, 8, 10, 14 from bottom to top. The red dot dashed lines are for $R_{\rm R}/M$=4.9, 4.93, 4.96, 4.99, 5.02, 5.05, 5.08, 5.11, 5.14, 5.17, 5.2, 5.23,
5.26, 5.29 from bottom to top. The blue dashed lines are for $R_{\rm T}/M$= 4.67, 4.71, 4.75, 4.79, 4.83, 4.87, 4.91, 4.95, 4.99, 5.03, 5.07, 5.11, 5.15, 5.19 from right to left. (a) Contour lines of $R_{\rm D}$ and $R_{\rm R}$. (b) Contour lines of $R_{\rm D}$ and $R_{\rm T}$. (c) Contour lines of $R_{\rm R}$ and $R_{\rm T}$.}\label{ppLRa}
\end{figure}

After obtaining the  curvature radii of points `D', `T', and `R', we depict in Fig. \ref{ppLRa} the respective contour lines $R_{\rm D}$, $R_{\rm T}$, and $R_{\rm R}$ in $\theta_{0}$-$a$ plane by using two of them. Each point in these figures is characterized by a pair values of ($a$, $\theta_{0}$). So if we  know any two  of $R_{\rm D}$, $R_{\rm T}$, and $R_{\rm R}$, we can read out ($a$, $\theta_{0}$) immediately from the figures.

\subsection{Approach III}

In the above two approaches, we showed how to use the curvature radius of different characteristic points to determine the black hole spin and the inclination angle. However in practice their variation as $a$ and $\theta_{0}$ change is very small. For a comparison, we list   in Table \ref{tab1} their values for $a$=0 and 1 for an equatorial observer. It is clear that the curvature radii $R_{\rm T}$ and $R_{\rm R}$ experience  small changes. The vertical diameter $\Delta\beta$ keeps the same value, while the horizontal diameter $\Delta\alpha$ and $R_{\rm D}$ significantly change.

\begin{figure}
\center{
\includegraphics[width=8cm]{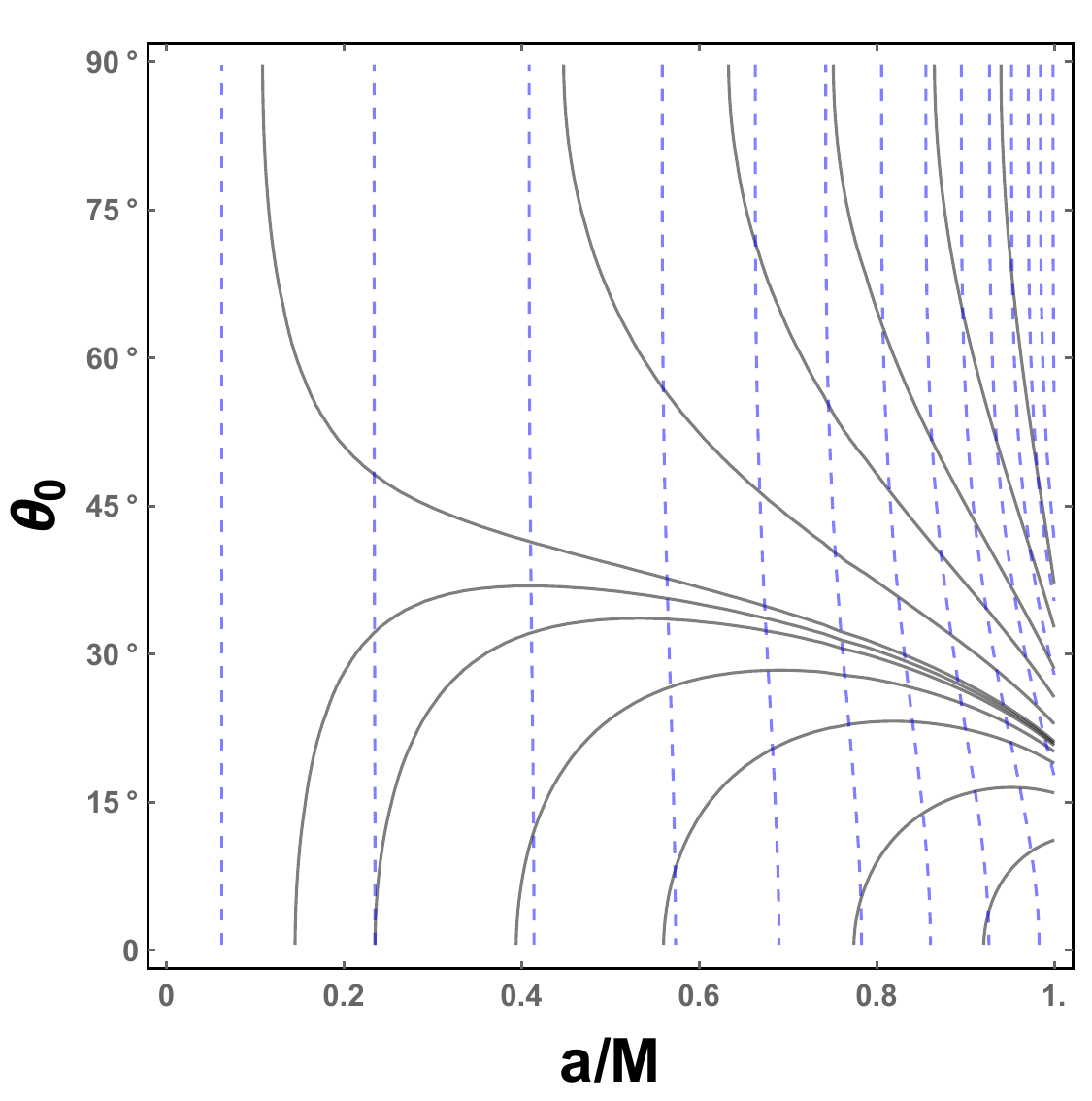}}
\caption{Contour lines of $\Delta\alpha$ and $R_{\rm D}$. The blue dashed lines are for $\Delta\alpha/M$=9.09, 9.29, 9.39, 9.49, 9.59, 9.69, 9.79, 9.89, 9.99, 10.09, 10.19, 10.29, 10.36, 10.39 from right to left. The black solid lines are for $R_{\rm D}/M$=4.9, 5, 5.1, 5.15, 5.18, 5.19, 5.2, 5.3, 5.5, 5.8, 6.5, 8, 10, 14 from bottom to top.}\label{ppTTAA}
\end{figure}

Consequently using $\Delta\alpha$ and $R_{\rm D}$ may yield  higher accuracy in determining $a$ and $\theta_{0}$.
Keeping this in mind, we show the contour lines of $\Delta\alpha$ and $R_{\rm D}$ in Fig. \ref{ppTTAA} with $\Delta\alpha/M$=9.09, 9.29, 9.39, 9.49, 9.59, 9.69, 9.79, 9.89, 9.99, 10.09, 10.19, 10.29, 10.36, 10.39, and $R_{\rm D}/M$=4.9, 5, 5.1, 5.15, 5.18, 5.19, 5.2, 5.3, 5.5, 5.8, 6.5, 8, 10, 14. Similarly, each point on the figure is characterized by a pair values of $(a, \theta_{0})$ and a pair values of $(\Delta\alpha, R_{\rm D})$. Hence the values of the parameters $a$ and $\theta_{0}$ can be easily read out by giving $\Delta\alpha$ and $R_{\rm D}$. Moreover, from Fig. \ref{ppTTAA}, we can find that, for fixed spin $a$, the horizontal diameter $\Delta\alpha$ almost keeps the same constant for different $\theta_{0}$.

\begin{table}[h]
\begin{center}
\begin{tabular}{cccccc}
  \hline\hline
 & $R_{\rm D}/M$ & $R_{\rm T}/M$ & $R_{\rm R}/M$ & $\Delta\alpha/M$ & $\Delta\beta/M$\\\hline
 $a/M=0$ & $3\sqrt{3}$ & $3\sqrt{3}$ & $3\sqrt{3}$ & $6\sqrt{3}$ & $6\sqrt{3}$\\
 $a/M=1$ & $\infty$ & $\frac{8}{\sqrt{3}}$ & $\frac{16}{3}$ & 9 & $6\sqrt{3}$\\\hline
 $\delta$ & $\infty$ & 0.5774 & 0.1372 & 1.3923 & 0\\
\hline\hline
\end{tabular}
\caption{Values for the curvature radii at points `D', `T', `R', the horizontal diameter $\Delta\alpha$, and the vertical diameter $\Delta\beta$ of the black hole shadow seen by an equatorial observer with $\theta_{0}=\frac{\pi}{2}$. The parameter $\delta$ denotes the change of these quantities between the Schwarzschild black hole $a/M=0$ and the extremal Kerr black hole $a/M=1$.}\label{tab1}
\end{center}
\end{table}

\section{Application to M87*}
\label{Application}

 A couple of days ago EHT showcased the first image of a black hole.  Specifically they obtained the image of the supermassive black hole located in the centre of the massive elliptical galaxy M87 \cite{EHT,EHT2}. This provides us with a new observational window to test strong gravity near the horizon of a black hole. Here we calculate the diameters and curvature radii for the black hole shadow of M87*. With  increased   observational precision we expect this can be helpful in testing the fine structure of   spacetime near a supermassive black hole.

Taking the values of  the parameters of the M87* from EHT observations \cite{EHT6},  i.e., the distance between us and M87* is $D=16.8$ Mpc and its mass is $M=6.5\times 10^{9}M_{\odot}$, we can calculate the gravitational radius \cite{EHT6}
\begin{equation}
 \theta_{g}=\frac{GM}{c^{2}D}\approx3.8 \;\mu \text{as} \quad .
\end{equation}
Since the approaching jet and the line of sight is $16^{\circ}$  \cite{Walker},
we only focus our attention on the viewing angle $\theta_{0}$ around $16^{\circ}$. Moreover, adopting the Standard and Normal Evolution (SANE) and Magnetically Arrested Disk (MAD) models, one can find from the rejection table of \cite{EHT2} that the black hole spin $|a|$=0.5 and 0.94 can pass these different constraints. Thus we consider values of $a$ within this range in our calculation.

\begin{figure}
\center{\subfigure[]{\label{M87alpha}
\includegraphics[width=8cm]{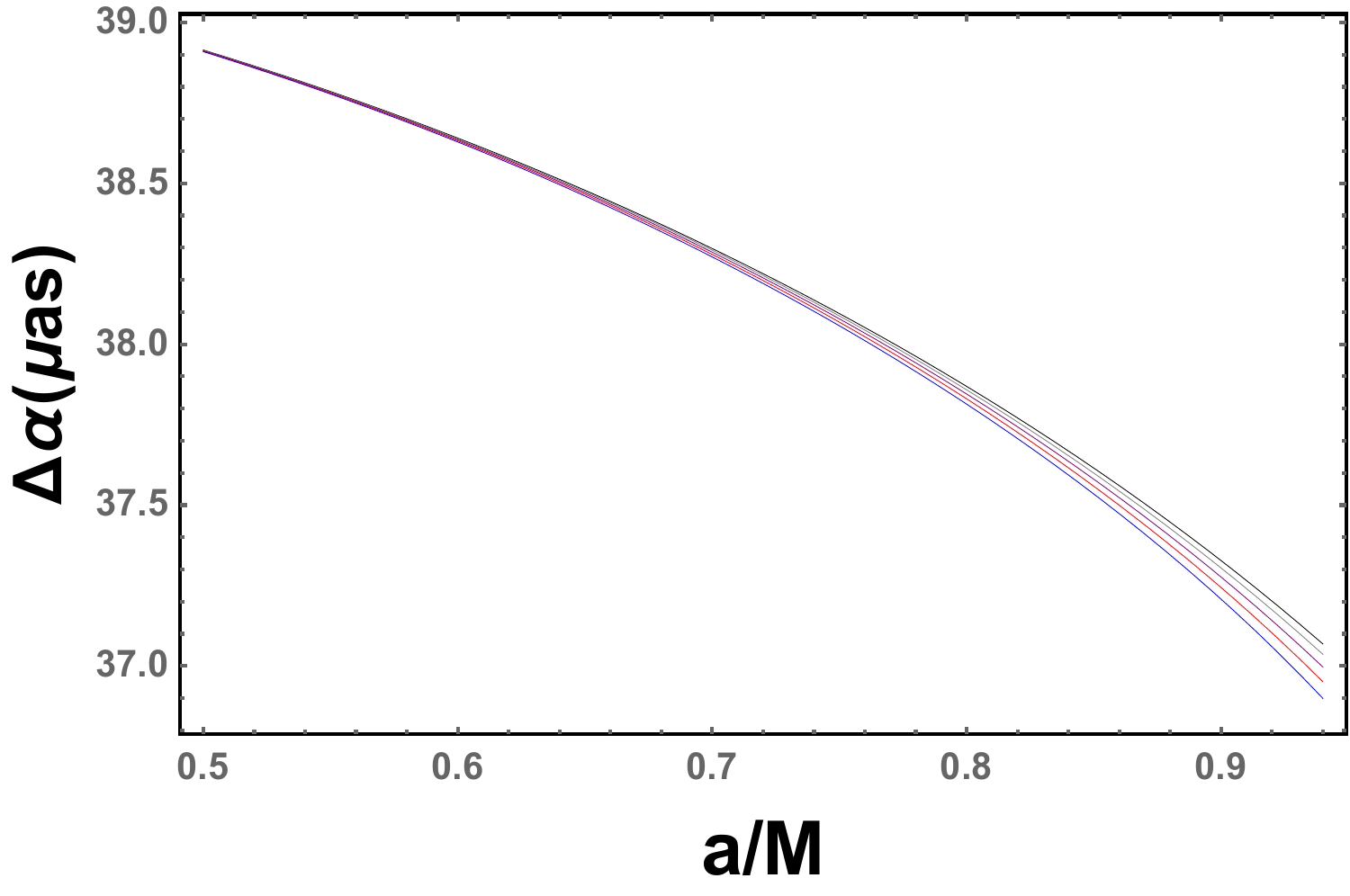}}
\subfigure[]{\label{M87beta}
\includegraphics[width=8cm]{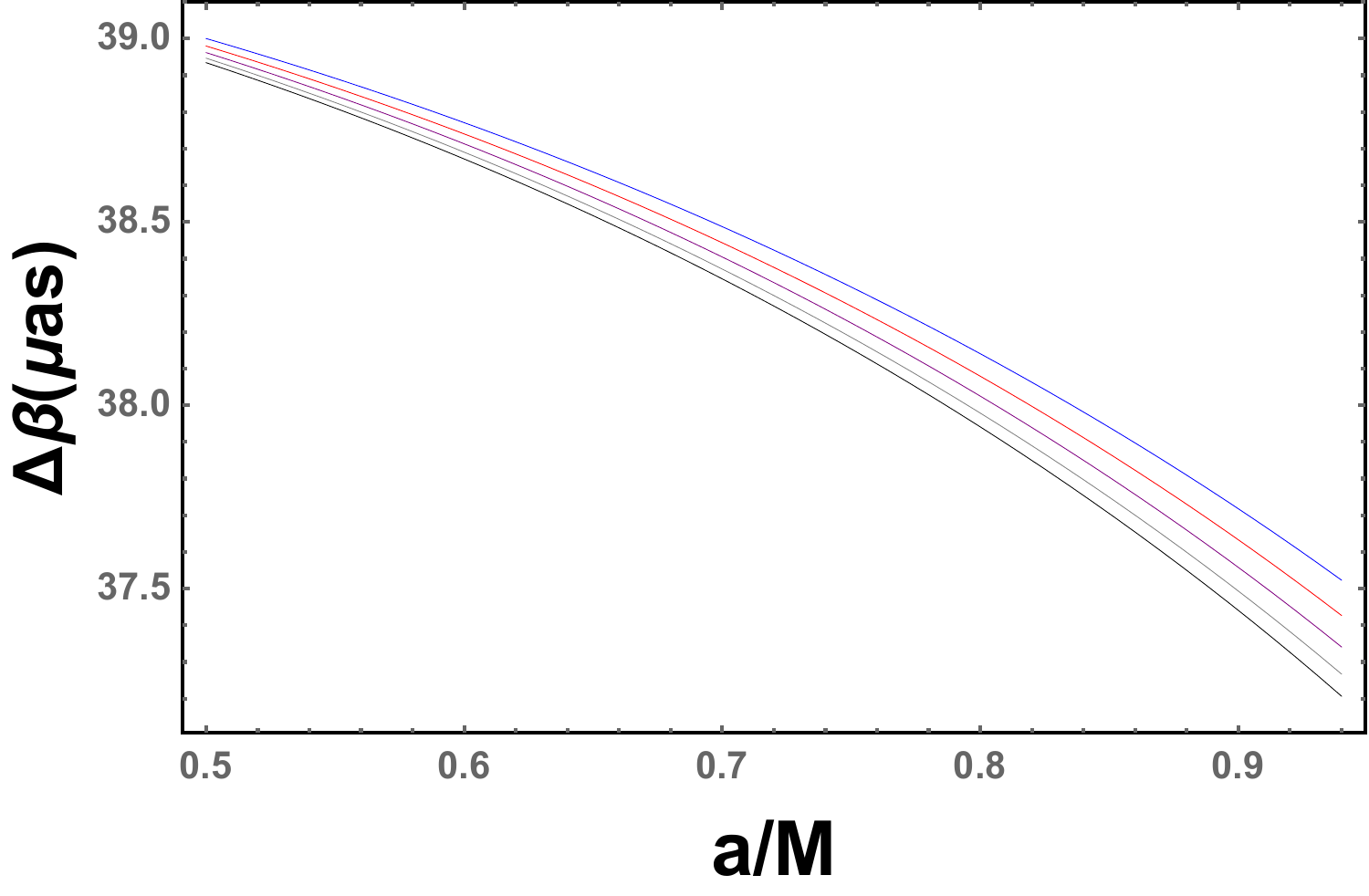}}}
\caption{The horizontal diameter $\Delta\alpha$ and the vertical diameter $\Delta\beta$ of the shadow cast by M87*. (a) $\Delta\alpha$ vs $a$. The viewing angle $\theta_{0}$=10$^{\circ}$, 13$^{\circ}$,16$^{\circ}$, 19$^{\circ}$, 22$^{\circ}$ from top to bottom. (b) $\Delta\beta$ vs $a$. The viewing angle $\theta_{0}$=10$^{\circ}$, 13$^{\circ}$,16$^{\circ}$, 19$^{\circ}$, 22$^{\circ}$ from bottom to top.}\label{pM87beta}
\end{figure}

 We show the horizontal   and   vertical diameters of the black hole shadow cast by M87* in Fig. \ref{pM87beta} with varying the viewing angle $\theta_{0}$=10$^{\circ}$, 13$^{\circ}$,16$^{\circ}$, 19$^{\circ}$, 22$^{\circ}$, respectively. For fixed $\theta_{0}$, both the diameters $\Delta\alpha$ and $\Delta\beta$ decrease with the black hole spin. For fixed black hole spin $a$, $\Delta\alpha$ decreases whereas $\Delta\beta$ increases with increasing $\theta_{0}$. At $a=0.5$, both $\Delta\alpha$ and $\Delta\beta$ are near 38.9 $\mu$as. When $a$ increases to 0.94, the vertical diameter $\Delta\beta$ is about 37.3 $\mu$as. The horizontal diameter $\Delta\alpha$ can even approach  37 $\mu$as. Thus the shape of the shadow will be increasingly deformed for high spin. Moreover, for low spin, $\Delta\alpha$ and $\Delta\beta$ are almost independent of the viewing angle $\theta_{0}$.

\begin{figure}
\center{\subfigure[]{\label{M87Rd}
\includegraphics[width=5cm]{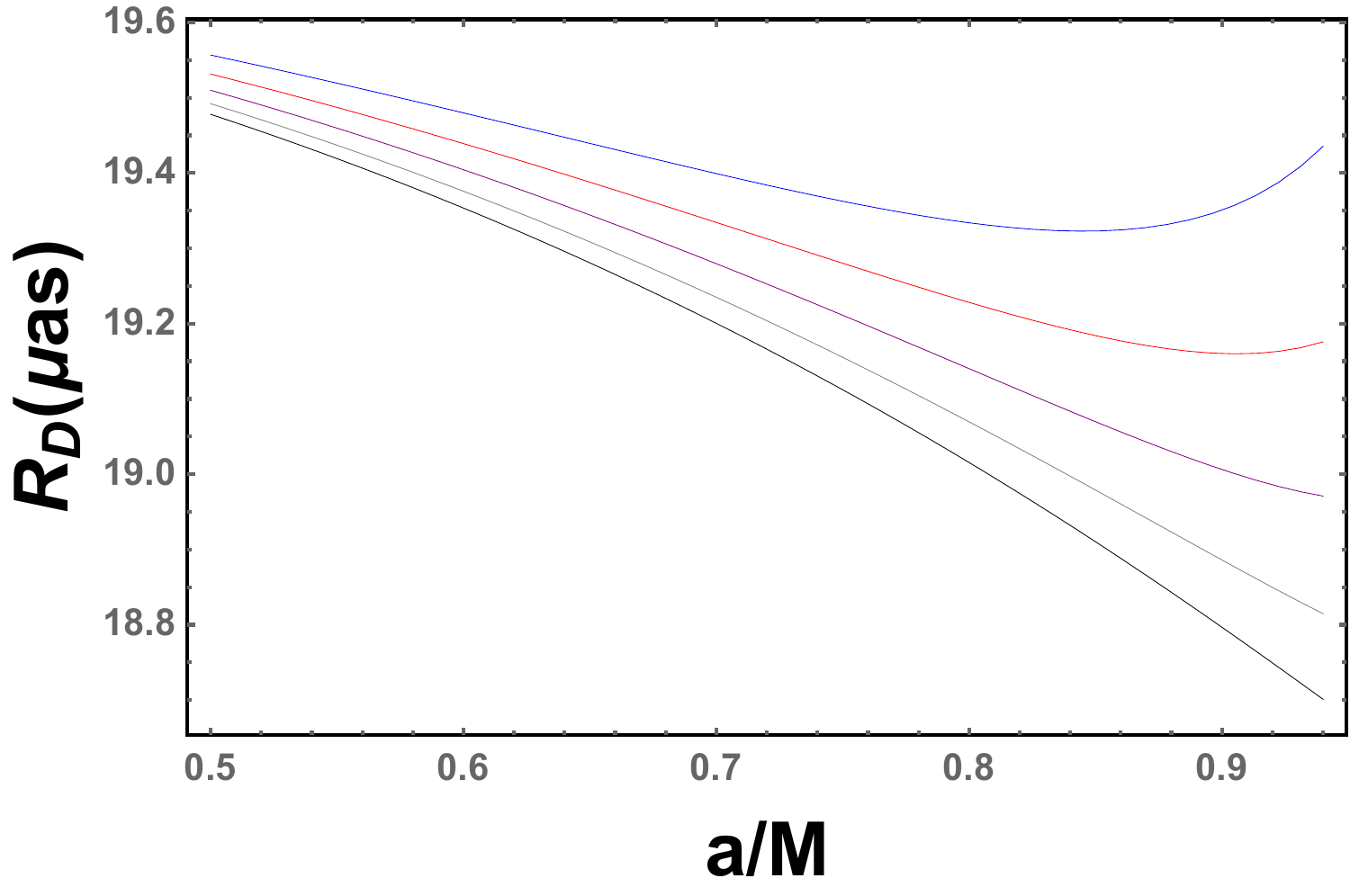}}
\subfigure[]{\label{M87Rr}
\includegraphics[width=5cm]{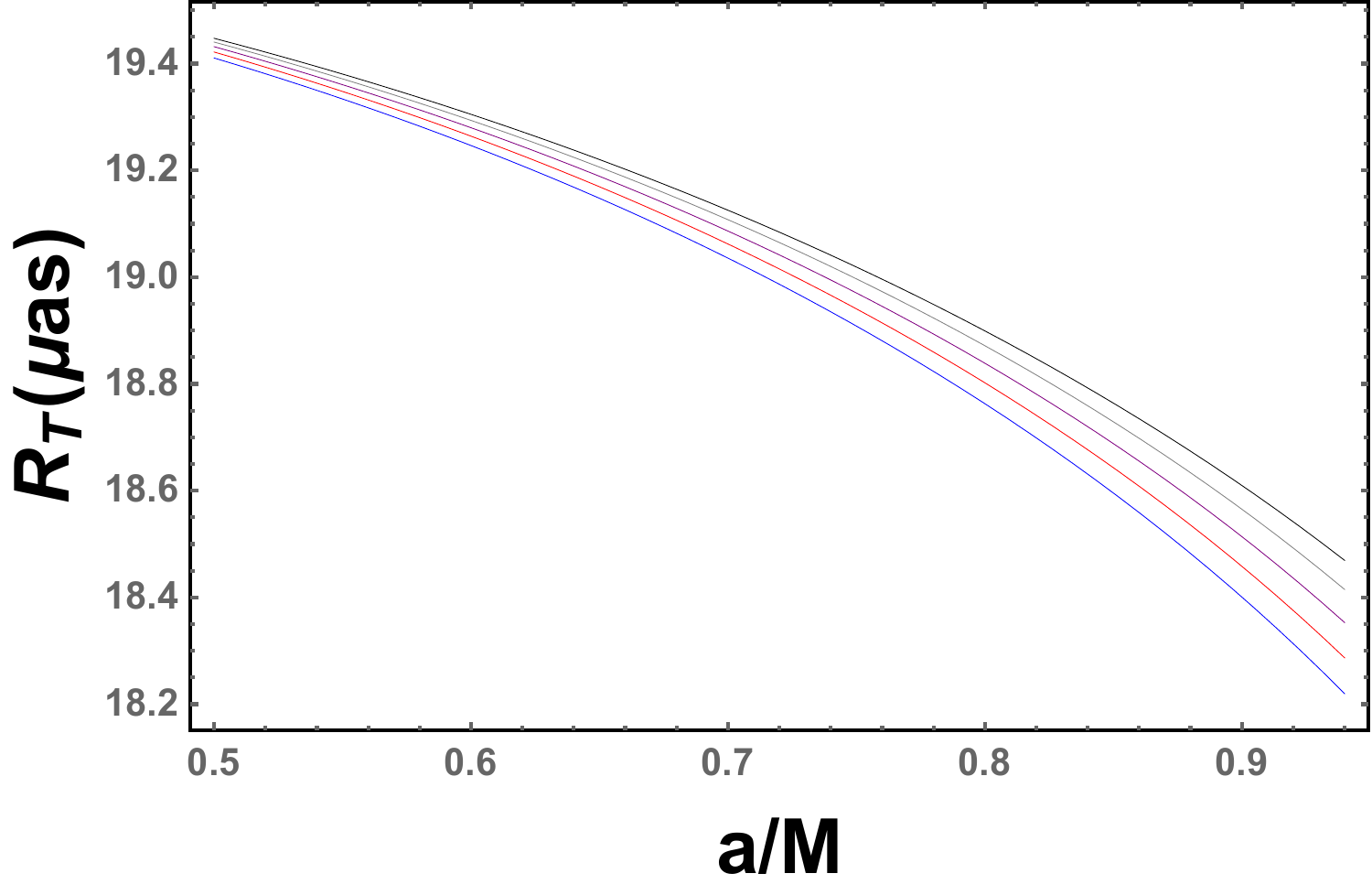}}
\subfigure[]{\label{M87Rt}
\includegraphics[width=5cm]{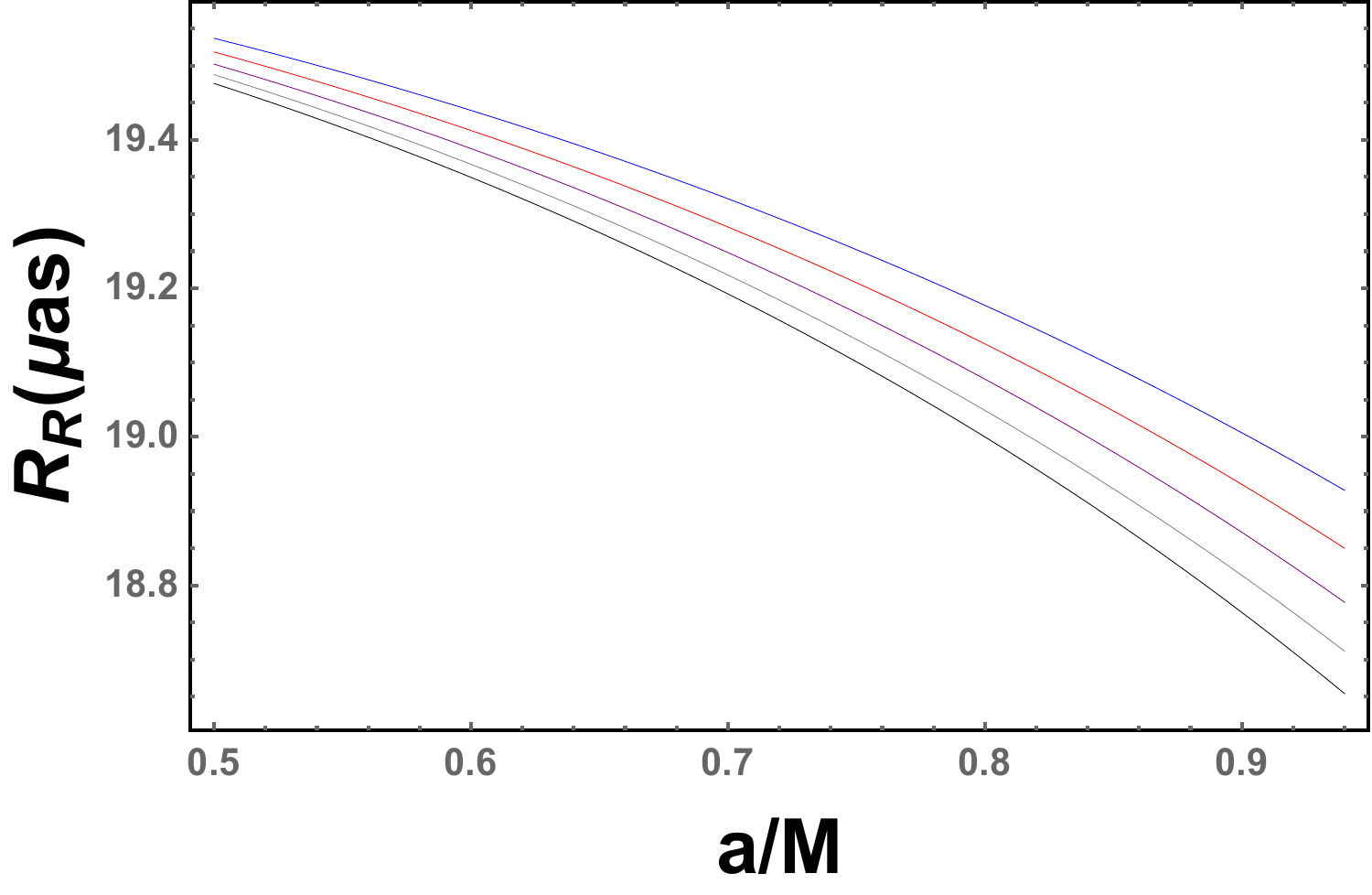}}}
\caption{Curvature radii at points `D', `T', `R' for the shadow cast by M87*. (a) $R_{\rm D}$ vs $a$. $\theta_{0}$=10$^{\circ}$, 13$^{\circ}$,16$^{\circ}$, 19$^{\circ}$, 22$^{\circ}$ from bottom to top. (b) $R_{\rm T}$ vs $a$. $\theta_{0}$=10$^{\circ}$, 13$^{\circ}$,16$^{\circ}$, 19$^{\circ}$, 22$^{\circ}$ from top to bottom. (c) $R_{\rm R}$ vs $a$. $\theta_{0}$=10$^{\circ}$, 13$^{\circ}$,16$^{\circ}$, 19$^{\circ}$, 22$^{\circ}$ from bottom to top.}\label{ppM87Rd}
\end{figure}

The curvature radii at points `D', `T', `R' are each plotted against the black hole spin in Fig. \ref{ppM87Rd} for $\theta_{0}$=10$^{\circ}$, 13$^{\circ}$,16$^{\circ}$, 19$^{\circ}$, and 22$^{\circ}$. For small viewing angle $\theta_{0}$, $R_{\rm D}$ decreases with the black hole spin. However for $\theta_{0}$   larger than 13$^\circ$, e.g., $\theta_{0}=22^{\circ}$, $R_{\rm D}$ first decreases from about 19.6 $\mu$as to 19.3 $\mu$as and then increases to 19.4 $\mu$as with increasing $a$. Thus, if $\theta_{0}>13^{\circ}$   we cannot  determine the black hole spin even if $R_{\rm D}$ is known, unless some other conditions are considered. Meanwhile, when $a$ varies from 0.5 to 0.94, $R_{\rm T}$ decreases from 19.4 $\mu$as to 18.2 $\mu$as, and $R_{\rm R}$ decreases from 19.5 $\mu$as to 18.7 $\mu$as.

\begin{table}[h]
\begin{center}
\begin{tabular}{ccccccc}
  \hline\hline
 $a/M$ & 0.5 & 0.6 & 0.7 & 0.8 & 0.9 & 0.94\\\hline
 $\Delta\alpha$($\mu$as)&38.91&38.64&38.29&37.85&37.28&37.00\\
 $\Delta\beta$($\mu$as)&38.96&38.71&38.40&38.03&37.56&37.34\\
 $R_{\rm D}$($\mu$as)&19.51&19.40&19.28&19.14&19.01&18.97\\
 $R_{\rm T}$($\mu$as)&19.50&19.39&19.25&19.08&18.87&18.78\\
 $R_{\rm R}$($\mu$as)&19.43&19.28&19.09&18.84&18.51&18.35\\\hline\hline
 \end{tabular}
\caption{Angular size of the quantities for the shadow cast by M87* with viewing angle $\theta_{0}=16^{\circ}$.}\label{tab2}
\end{center}
\end{table}

For comparison, we list the values of these quantities $\Delta\alpha$, $\Delta\beta$, $R_{\rm D}$, $R_{\rm T}$, and $R_{\rm R}$ in Table \ref{tab2} when $\theta_{0}=16^{\circ}$ is fixed. For two small values of the spin, the differences between them are very tiny. However, for two high spins $a/M$=0.7 and 0.94, the differences of $\Delta\alpha$, $\Delta\beta$, $R_{\rm D}$, $R_{\rm T}$, and $R_{\rm R}$ between them are, respectively, about 1.3, 1.1, 0.3, 0.5, and 0.7 $\mu$as. Thus, when the precision of EHT is improved to around 1 $\mu$as comparing with current resolution 5 $\mu$as in the final reconstructed images, the black hole spin can be well determined in high spin case.

\section{Conclusions and discussions}
\label{Conclusion}

In this paper, we mainly considered the application of the curvature radius we previously proposed \cite{WeiLiu}. Based on the characteristic points and making use of the curvature radius, we have put forward three novel approaches to determine the spin of the Kerr black hole and the inclination angle of the observer.

The first approach makes use of only two symmetric characteristic points, `T' and `B' to determine $a$ and $\theta_{0}$.  This is the fewest number of points required, and is in contrast to previous approaches  \cite{Hioki} requiring four characteristic points, `D', `R', `T', and `B'.

In the second approach we obtain $a$ and $\theta_{0}$ by only finding the curvature radii of these characteristic points.

The third approach employs the horizontal diameter $\Delta\alpha$ and the curvature radius $R_{\rm D}$ of point `D' to determine $a$ and $\theta_{0}$.   The former quantities have the most sensitive dependence on the latter and so may provide a more accurate way to determine the spin and the viewing angle.

The three approaches above can provide an  analytic result from a rather small amount of information.  In  astronomical observations \cite{EHT,EHT2,Hughes2015,Sanders2013}   shadows are expected to be obtained with uncertainty, and a fit  to the observed image might be performed to obtain the black hole parameters usin Eqs. (\ref{oe}) and (\ref{oe2}).

Although our results are mainly for the Kerr black hole, it is easy to extend to other black hole backgrounds, where perhaps, more black hole parameters are present.  Consequently more characteristic points will have to be considered to fix the black hole parameters and the inclination angle of the observer. Though it remains to be demonstrated, we believe that our approaches will still be applicable to determine the various black hole parameters.

Finally, we applied our study of the curvature radius to the supermassive  black hole M87*. We computed the angular size for the horizontal diameter $\Delta\alpha$, the vertical diameter $\Delta\beta$, and curvature radii $R_{\rm D}$, $R_{\rm T}$, $R_{\rm R}$. When the black hole spin increases from 0.5 to 0.94, both $\Delta\alpha$ and $\Delta\beta$ are about 38.9 $\mu$as, and then approach   37.0 $\mu$as and 37.3 $\mu$as, respectively. For $a/M=0.5$, the curvature radii $R_{\rm D}$, $R_{\rm T}$, and $R_{\rm R}$ are all about 19.5 $\mu$as, and then go to 19.0 $\mu$as, 18.8 $\mu$as, and 18.4 $\mu$as, respectively. With improved observational resolution  and reconstruction, we expect that the nature of M87* will be well determined through its shadow.

In closing, by focussing on the local properties of the shadow at several characteristic points, we have shown that   information about the black hole can be completely determined from these local properties. However for an astronomical applications, this requires obtaining a sufficiently high resolution image of the shadow edge. Although the current  resolution of the EHT cannot achieve this,  we expect that our method will applicable in the (near) future once high-resolution shadow images become available. In contrast to integrated approaches that combine information over the entire shadow edge, our method provides a useful alternative for extracting information from the shadows of black holes. Furthermore, there exist  bright lensing rings  around black hole shadows, and these are more easily observed. The innermost one is very close to the black hole edge, and it can be naturally treated as if it were the black hole edge, in which case  our approach can be easily used. Moreover, we can directly apply our concept of the curvature radius, since the light ring is also one dimensional, and  then extract information about the black hole.

In summary, the curvature radius is a novel concept, one that we expect will prove useful from both theoretical and observational perspectives in understanding the strong gravitational effects of black holes.

\section*{Acknowledgements}
This work was supported by the National Natural Science Foundation of China (Grants Nos. 11675064, 11875151, 11522541, and U1738132) and the Natural Sciences and Engineering Research Council of Canada. S.-W. Wei was also supported by the Chinese Scholarship Council (CSC) Scholarship (201806185016) to visit the University of Waterloo.

\end{document}